# Blind Source Separation in Biomedical Signals Using Variational Methods

Yasaman Torabi[1], Shahram Shirani[1,2], James P. Reilly[1]

[1]Electrical and Computer Engineering Department, McMaster University, Hamilton, Canada  [2] L.R. Wilson/Bell Canada Chair in Data Communications, Hamilton, Canada

CORRESPONDING AUTHOR: Yasaman Torabi (e-mail: torabiy@mcmaster.ca).

## Abstract

This study introduces a novel unsupervised approach for separating overlapping heart and lung sounds using variational autoencoders (VAEs). In clinical settings, these sounds often interfere with each other, making manual separation difficult and error-prone. The proposed model learns to encode mixed signals into a structured latent space and reconstructs the individual components using a probabilistic decoder, all without requiring labeled data or prior knowledge of source characteristics. We apply this method to real recordings obtained from a clinical manikin using a digital stethoscope. Results demonstrate distinct latent clusters corresponding to heart and lung sources, as well as accurate reconstructions that preserve key spectral features of the original signals. The approach offers a robust and interpretable solution for blind source separation and has potential applications in portable diagnostic tools and intelligent stethoscope systems.

## Audiovisual Material

A video presentation of this work is available at:

https://youtu.be/p2A_qRGljIc?si=ZS9JpOQCPeDi1ZPS

## Introduction

Analyzing heart and lung sounds is critical in clinical diagnostics, yet these signals often overlap, making it difficult to isolate and interpret each source accurately. Manual separation requires expert skill and remains error-prone, especially in noisy environments. This work addresses the challenge using an unsupervised approach based on Variational Autoencoders (VAEs), a type of deep generative model that learns to encode mixed signals into a structured latent space. Unlike traditional source separation techniques, this model operates without prior knowledge about the nature or number of sources and requires no labeled data.

Several techniques have been developed for blind source separation (BSS), including non-negative matrix factorization (NMF) approaches that exploit temporal and spectral patterns in cardiorespiratory signals [1,2]. While effective, these methods often lack the generative flexibility of VAEs, which can better model complex, overlapping acoustic phenomena [3]. Recent advancements in digital stethoscope design and wearable biosensors have enabled high-quality acquisition of cardiopulmonary signals, providing rich data for training deep models [4-6]. Deep neural networks—including CNNs, RNNs, and hybrid architectures—have shown promise in a wide range of biomedical applications, from chronic pain assessment [7] and fever detection [8] to ICU mortality prediction [9] and ventilator-associated pneumonia risk [10]. Not only cardiovascular applications, but also fields like neuroscience modeling, benefit from unsupervised frameworks [11,12]. These developments can be further supported by network architecture optimization strategies for more efficient and robust processing [13-16].

Our proposed variational method was applied to real cardiopulmonary recordings obtained from a clinical manikin using a digital stethoscope. The VAE architecture learns a compact latent representation, enabling the decoder to reconstruct individual components by sampling from distinct latent regions. During training, distinct clusters emerged in the latent space, corresponding to heart and lung sources, as confirmed by t-SNE visualizations. Spectrograms of the reconstructed signals closely matched the ground truth components, demonstrating successful separation. This work not only highlights the potential of VAEs in biomedical audio analysis but also contributes to broader numerical strategies for solving inverse problems in scientific computing.

## Methods

We collected mixed heart and lung sound recordings from a clinical manikin using a 3M Littmann digital stethoscope, ensuring a controlled acquisition environment [17]. These recordings served as input to a Variational Autoencoder (VAE) comprising an encoder–decoder architecture. The encoder learns a variational posterior $q_\phi(z \mid x)$ over latent variables $z$, while the decoder models the likelihood $p_\theta(x \mid z)$. The model is trained by maximizing the evidence lower bound (ELBO), which balances reconstruction accuracy and latent space regularization using Kullback–Leibler divergence.

Let $x \in \mathbb{R}^n$ denote the observed mixed signal, assumed to be generated from a set of latent variables $z \in \mathbb{R}^k$ such that $x \sim p_\theta(x|z)$, where $p_\theta$ is a decoder parameterized by neural network weights $\theta$. Direct computation of the posterior $p_\theta(z|x)$ is intractable, so we approximate it with a variational distribution $q_\phi(z|x)$, parameterized by an encoder with weights $\phi$. The model is trained by maximizing the Evidence Lower Bound (ELBO):

$$L(\theta, \phi; x) = \mathbb{E}_{q_\phi(Z|X)}[\log p_\theta(x|z)] - D_{KL}\left(q_\phi(z|x) || p(z)\right).$$

where $D_{KL}$ denotes the Kullback–Leibler divergence and $p(z)$ is a standard normal prior. The first term encourages accurate reconstruction, while the second regularizes the latent space. The separated signals are obtained by decoding distinct regions in the latent space associated with different acoustic sources. We visualized the latent space using t-SNE to assess whether the model captured meaningful structure for source separation. Performance was evaluated by comparing the time-frequency spectrograms of the reconstructed signals against the ground truth components.

## Results

The trained VAE model effectively separated heart and lung components from the input mixtures. Visualization of the latent space using t-SNE revealed the emergence of distinct clusters corresponding to the two sound sources as training progressed. These clusters indicate that the model successfully encoded source-specific features in the latent representation. The decoder was able to reconstruct the individual signals with high fidelity, preserving the key spectral patterns of each source. Time-frequency spectrograms of the predicted components showed strong alignment with the true signals, confirming the model's ability to perform blind source separation without supervision.

Figure 1 shows the t-SNE visualization of the latent space learned by the model. Over the course of training, distinct clusters emerge that correspond to heart and lung sound components, indicating that the model has successfully captured meaningful structure for separating the sources. Figure 2 displays the time-frequency representations of the predicted signals compared to the original mixtures. These spectrograms demonstrate the model's ability to reconstruct source-specific frequency patterns.

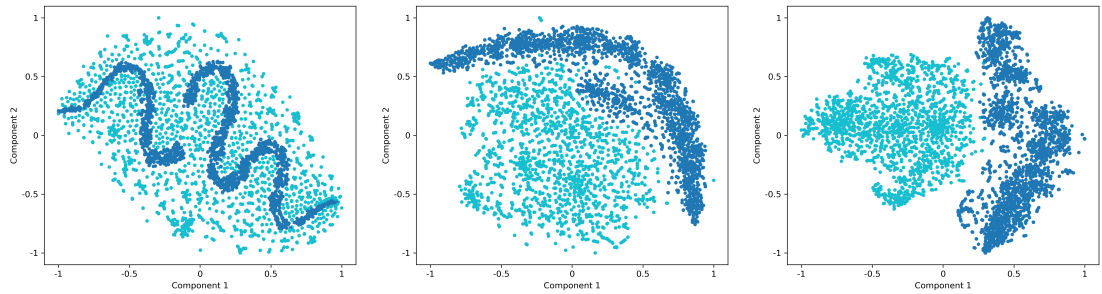

**Figure 1.** t-SNE projections of the latent space at successive training epochs. As training progresses (left to right), the latent representations of heart and lung sounds begin to separate into distinct clusters.

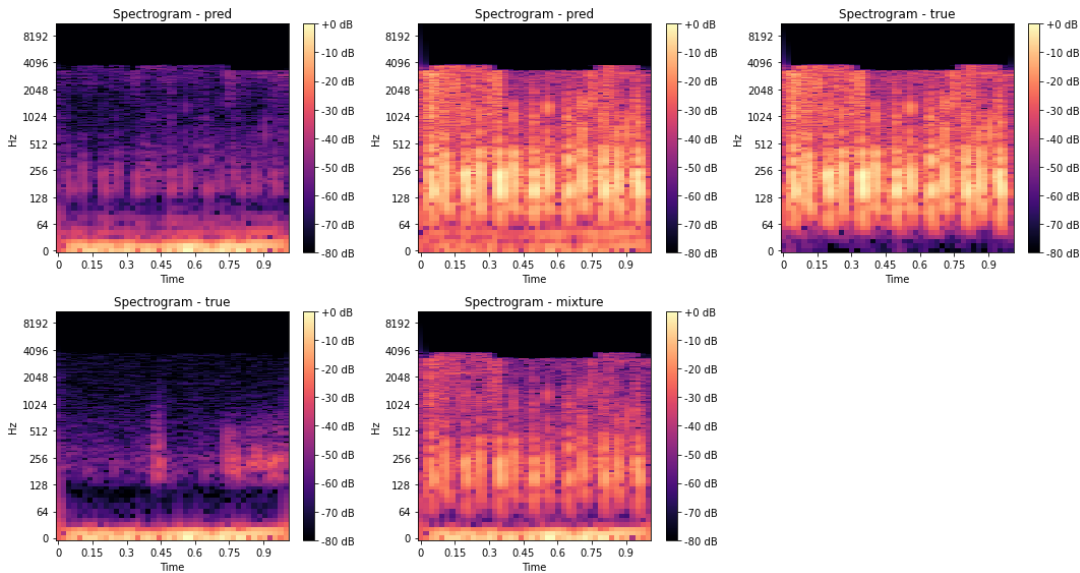

**Figure 2.** Spectrogram comparison of predicted and true sources. The top row shows the predicted spectrograms for individual components, while the bottom row includes the true sources and their corresponding mixture. The predicted signals preserve key spectral features of the original heart and lung sounds.

## Discussion

This study demonstrates that a variational autoencoder can learn to separate overlapping heart and lung sounds in an unsupervised manner by exploiting the structure of the latent space. The formation of distinct latent clusters and the preservation of spectral features in the reconstructions suggest that the model captures meaningful representations of the underlying sources. Compared to traditional autoencoders or supervised methods, the VAE offers improved interpretability, flexibility, and generalization, particularly in settings where labeled data are unavailable. The results also highlight the potential for integrating such models into portable diagnostic tools, enhancing real-time auscultation and clinical decision support.